\documentclass[journal,comsoc]{IEEEtran}

\usepackage{amssymb,color}
\usepackage{amsmath}
\usepackage{bbm}
\usepackage{graphicx}
\usepackage{epstopdf}
\usepackage{amsthm,amsfonts}
\usepackage{subcaption}

\usepackage{accents}

\usepackage{algorithm}
\usepackage{algorithmic}
\usepackage{psfrag}
\usepackage{cite}
\usepackage{pifont}
\usepackage[font={small}]{caption}

\allowdisplaybreaks

\makeatletter
\newcommand{\oset}[3][0ex]{%
  \mathrel{\mathop{#3}\limits^{
    \vbox to#1{\kern-2\ex@
    \hbox{$\scriptstyle#2$}\vss}}}}
\makeatother

\begin{document}
\title{On Clustering and Channel Disparity in Non-Orthogonal Multiple Access (NOMA)}

\author{
\IEEEauthorblockN{\large  Konpal Shaukat Ali$^{\dagger}$, Mohamed-Slim Alouini$^{\ast}$, Ekram Hossain$^{\dagger}$, and Md. Jahangir Hossain$^{+}$}

\thanks{
$^{\dagger}$ The authors are with the Department of Electrical and Computer Engineering, University of Manitoba, Winnipeg, Canada (Email: \{konpal.ali, ekram.hossain\}@umanitoba.ca).

$^*$The author is with the Computer, Electrical, and Mathematical Sciences and Engineering (CEMSE) Division, King Abdullah University of Science and Technology (KAUST), Thuwal, Makkah Province, Saudi Arabia (Email: slim.alouini@kaust.edu.sa).

$^+$The author is with the School of Engineering, the University of British Columbia (Okanagan Campus), Canada (Email:  jahangir.hossain@ubc.ca). 

 }}


%
%
%

\maketitle

\begin{abstract} 
Non-orthogonal multiple access (NOMA) allows multiple users to share a time-frequency resource block by using  different power levels. An important challenge associated with NOMA is the selection of users that share a resource block. This is referred to as clustering, which generally exploits the channel disparity (i.e. distinctness) among the users. We discuss clustering and the related resource allocation challenges (e.g. power allocation) associated with NOMA and highlight open problems that require further investigation. We review the related literature on exploiting channel disparity for clustering and resource allocation. There have been several misconceptions regarding NOMA clustering including: 1) clustering users with low channel disparity is detrimental, 2) similar power allocation is disastrous for NOMA. We clarify such misunderstandings with numerical examples. 



\end{abstract}


\section*{The Concept of Multiple Access}
\subsection*{Orthogonal Multiple Access (OMA)}
Wireless communication as we know it is {confined by} two fundamental resources: time and spectrum. 
Each generation of cellular communication has made efforts to improve how these resources are managed to improve the \emph{amount} of communication that can be done given a set of these resources. This amount is limited by the interference the communicating entities will cause to one another. A natural way to deal with this was to split the resources in such a way that users accessed them in an orthogonal fashion thereby avoiding interference with one another; this is the well known concept of orthogonal multiple access (OMA). In the second generation (2G) cellular network, this was achieved by time division multiple access (TDMA). TDMA involves serving different users in different time slots, where a user would have access to the entire frequency channel in its time slot. Another OMA technique is frequency division multiple access (FDMA) which allows a user to be able to communicate at any time (i.e. full access to the time resources) by giving each user a slot of frequency rather than access to the entire frequency resource. While TDMA restricts a user to communicating in a particular time slot only, FDMA allows the freedom of having the ability to communicate at any time at the price of reduced spectrum resources. With the fourth generation (4G) a more efficient technique, namely orthogonal frequency division multiple access (OFDMA) was deployed which splits both the time and frequency resources into a grid so that a user can be assigned different blocks in the grid, known as time-frequency resource blocks. This allows significantly more flexibility in resource management by assigning time-frequency resource blocks to users depending on 1) when they need to communicate, 2) how much communication is required. 



\subsection*{Non-Orthogonal Multiple Access (NOMA)}

With the growing demands in data rate and ever-increasing traffic,  non-orthogonal multiple access (NOMA) has emerged as a promising technique for the fifth generation (5G) and beyond cellular networks \cite{N1,N8}. NOMA allows multiple users to share a time-frequency resource block by superposing the messages in either the power domain or the code domain. We focus on power domain NOMA in this article which allows multiple users to share a time-frequency resource block by allocating different power levels for the message of each user. This in contrast to OMA where a single user would have access to the full power resources in its time-frequency resource block. The price paid for the increased communication that NOMA allows is intracell interference between the multiple users that share time-frequency resources. NOMA employs successive interference cancellation (SuIC) techniques to mitigate the impact of and manage this intracell interference, so that that the messages of interest of each user can be successfully retrieved. 

In the context of NOMA, let us define the following terminologies:
\begin{itemize}
\item Cluster: The set of users that share a time-frequency resource block and communicate via NOMA over these resources, interfering with one another in the process.
\item Clustering: The process of selecting which users are included in a particular cluster. In two-user NOMA this is often referred to as `pairing'.
\end{itemize} 

By carefully allocating resources to the users in a cluster, NOMA allows us to meet very specific network requirements. While the time-frequency resource block is a restrictive grid in terms of what can be reaped via OMA, modifications in the continuous power dimension via NOMA gives more flexibility. This flexibility will be valuable in next generation networks with their plethora of devices with varying requirements from ultra high reliability and low latency communication (URLLC) to low quality of service (QoS) requirements in certain internet of things (IoT) devices, and everything in between. Typically in NOMA, resources such as power are not allocated to the users in an equal or seemingly fair manner. In fact, the concept of SuIC requires power allocation such that decoding some of the messages is easier than decoding others. SuIC  aims to successively decode messages in the chain. The decoded message is then subtracted from the remaining signal, and the next message in the chain is decoded. Accordingly, resources are allocated such that messages that need to be decoded first in the chain are easier to decode since decoding each subsequent message depends on the success of decoding the preceding messages. 



\subsubsection{Downlink NOMA}{In the downlink, resources are allocated such that decoding the message of a weaker user is easier than decoding the message of a stronger user. In the two-user NOMA scenario, this can be done by allocating messages of the weaker user higher power and/or lower target rate, and vice versa for the stronger user. This way the weak user with its poor channel will be aided in decoding its message easily. The strong user on the other hand, will first decode the message of the weaker user, subtract it from the signal and then decode its own message. Although the message of the strong user is harder to decode, since it has a good channel, it will be able to afford doing this. This concept is extended to the scenario where more than two users share one time-frequency resource block; each user decodes the messages of all weaker users and treats the messages of all stronger users as noise.}

\subsubsection{Uplink NOMA} In the uplink, on the other hand, decoding takes place at one receiver, i.e. the base station (BS). Since SuIC entails decoding the easier-to-decode message first\footnote{While decoding the easiest message first is the smart strategy, it is not mandatory. In fact, SuIC in both the uplink and downlink can be done in any random order; however, the performance quality would suffer.}, simple fixed power transmissions by the users could be used. The BS would begin by decoding the message of the strongest user first in this case, as it would have the best signal due to its best channel. Then messages of the remaining users would be successively decoded in order of their channel and consequently signal strength, with the weakest user's message being decoded last. 




Hence, due to SuIC, decoding the message of a user in NOMA, unlike OMA, becomes a joint event. Clearly, employing SuIC implies there is a need to rank users in terms of some measure of channel conditions or strength; {this is referred to as user ordering in NOMA}. There are multiple ways to decide what makes a user stronger or weaker than another. A technique frequently used in the literature is ordering users based on the distance to the serving BS, i.e. link distance \cite{myNOMA_icc,myNOMA_tcom,N2}. Other techniques take into account variables such as fading, noise and intercell interference\cite{myNOMA_tcom,N8,N9,N15}. 

\subsection*{Going from Single-Cell to Multi-Cell NOMA: Large-Scale NOMA}
The idea behind employing OMA was to completely mitigate the interference between the users being served by a BS. However,  interference from the rest of the network, i.e. outside of the cell of interest, still exists; this is referred to as intercell interference. With the increased network densification, intercell interference will be a dominant source of interference for NOMA, despite the presence of intracell interference which is inherent to NOMA \cite{my_nomaMag}. 
Therefore, it is crucial to study the impact of intercell interference in the evaluation of large-scale NOMA systems and develop methods to mitigate the intercell interference. Techniques such as stochastic geometry tools have been shown to be instrumental to analyze large-scale NOMA systems \cite{myNOMA_tcom,N6,N18,myNOMA_meta,myNOMA_icc,N16}.


\section*{Challenges Associated With NOMA}

Some of the fundamental challenges associated  with NOMA include: determining the user clustering, cluster size, user ordering, cluster objective, and resource allocation. {It is important to note that these challenges may be intertwined. Additionally, there is no particular order in which these challenges need to be addressed for a NOMA setup, and the order in which they are addressed impacts the decisions made. Since this article focuses on clustering, we highlight the impact of clustering on each of the other challenges.} 

\subsection*{User Clustering}
{Two challenges associated with user clustering are}: 1) choosing the area(s) from within which users for a cluster will be selected, i.e. spatial restriction on NOMA operation 2) channel disparity restrictions, if any, between the users in a cluster. {Note that since NOMA requires sharing a resource block with users served by the same BS, selecting users from all over the Voronoi cell is an intrinsic spatial restriction. Of course more stringent restrictions can be applied.}


Despite its importance, user clustering in NOMA has been a quite lightly touched subject and there are not many explicit studies on this in the literature particularly in the context of large networks. Often works on NOMA attempt to address both of these factors by using simple clustering techniques such as a disk and an annulus surrounding it with fixed radii for the two-user NOMA case \cite{N4}. {This in a way imposes a spatial restriction on the users and if the annulus and central disk do not share a boundary, introduces a minimum channel disparity in terms of a minimum distance between the users in the two areas.} Other techniques to address the first challenge include the commonly used approach of having users distributed uniformly in the cell and selecting users in a cluster arbitrarily at random \cite{N16}. In \cite{N6}, the authors select NOMA users by selecting users from within a disk of a fixed radius around each BS; however, this risks not having nearest BS association. In \cite{myNOMA_tcom,myNOMA_icc}, the authors introduce a more sophisticated clustering technique which restricts selecting NOMA users from a disk around the BS where the size of the disk varies with intercell interference conditions, guaranteeing that it lies inside the Voronoi cell. In \cite{myNOMA_tcom}, three schemes that restrict clustering to improve interference conditions are shown to further improve network performance. The works in \cite{myNOMA_tcom,myNOMA_icc,N16,N6} use ordered distance or large-scale channel strength distributions to account for the impact of variations in channel disparity and study average network performance of large networks. Hence, although they take into account channel disparity in their assessment, they do not address the second challenge related to user clustering explicitly. In \cite{N18}, two-user downlink NOMA is studied and the second challenge is addressed by studying a clustering technique where users are selected such that the strong user has SINR above threshold $T_1$, while the weak user has SINR below threshold $T_2$, where $T_1 \geq T_2$. This ensures a certain minimum channel disparity in terms of the difference $T_1-T_2$ between the SINRs of the two users.

{In regards to the channel disparity restrictions associated with the user clustering process, it is important to emphasize that quantifying channel disparity is not trivial. In fact, it is even complicated to define what the true channel disparity is; it needs to account for the difference in channel strength and conditions of the users sharing a resource block. It should be noted that the channel disparity measure does not have to be based on the same measure of channel strength being used for user ordering. Additionally, more often than not, the impact of spatial restrictions and channel disparity can be intertwined. 
This is an open area of research which is still not very well defined.}

\subsection*{Cluster Size}
A number of works on NOMA focus on the simple two-user NOMA case where only two users share a time-frequency resource block. This may not necessarily be the optimum approach. NOMA is promoted for a number of services such as IoT where the data rate requirements may not be too high but a very large number of devices may need to be served. In such a situation, a larger number of users can exploit the time-frequency resource block and meet their data rate requirements. In \cite{myNOMA_tcom}, the existence of optimum cluster sizes that maximize the sum-rate of the cluster subject to both a QoS requirement and a symmetric rate requirement are shown. Additionally, it is shown that the available resources are able to support a maximum cluster size given the requirements. In a more flexible setting, the maximum cluster size that can be supported as well as the optimum cluster size would change depending on the network requirements and goals. It is an important area to explore particularly in light of the increase in demand of low latency communication as well as communication with {variable reliability requirements}. Hence, determining the cluster size not only depends on the network conditions such as interference and SuIC efficiency, but needs to take into account the network objective. Also, note that, for clustering, the spatial and channel disparity restrictions can reduce the number of available users to select from and thus can result in the selection of smaller clusters. Currently, there is plenty of room to investigate the interplay between clustering and cluster size. 





\subsection*{User Ordering}
As has been mentioned earlier, SuIC requires ranking of users in some order, generally based on some measure of channel strength. In theory, the users in a cluster can be ordered arbitrarily at random and resource allocation followed by SuIC can be done based on this order; however, such ordering will most likely result in severely compromised performance. This is because SuIC is intended to exploit the channel difference in such a way that in the downlink users with good channels do more ``work'' decoding their message than users with worse channels. Similarly, in the uplink the BS does more work in decoding the messages coming from users with weaker channels than users with better channels. Hence, ordering users based on a sensible measure of channel strength is important for enhancing the performance of SuIC. 

As has been mentioned in the previous section, one of the most common user ordering techniques is based on the link distance. Although simple, it is quite an effective technique since this distance has one of the most significant impacts on the signal portion of the SINR and consequently on performance. Other ordering techniques that take into account variables such as fading, intercell interference, and noise have also been used in the literature. One technique that incorporates more information than the link distance based is ranking according to the total received signal-to-intercell-interference-and-noise ratio. Although one may be inclined to think that an ordering technique that incorporates more channel information would always result in superior performance, it is shown in \cite{myNOMA_tcom} that this is not always the case due to the nature of SuIC. While there are areas of the rate region, where the latter outperformed the former in terms of performance, the link distance-based ordering was superior in other areas. The existence of an optimum ordering technique is an open problem. In fact, it is currently not even known if there exists an optimum ordering technique, or if this varies with network conditions and objectives. 

{It should be noted that if the order of the users is known ahead of clustering, the knowledge of the rank and quality of the users can be used in the clustering process. For instance, the measure of channel strength can be used to put channel disparity restrictions on the NOMA users to be selected. Similarly, quality-based restrictions can be placed on the users to be selected in a cluster. The impact of user ordering on clustering is an interesting direction for future work.}

\subsection*{Cluster Objective}
Defining an explicit cluster objective naturally depends on the goal of the network or cluster but it can also vary with the network conditions. Often a number of goals with varying flexibility are anticipated to be met. In such a situation, optimizing the cluster objective to meet these requirements in the best possible way can become complex. 
Since addressing each challenge associated with NOMA does not have to be done in a particular order, it becomes more complicated to decide, for instance, if the cluster objective should be decided after selecting a cluster size or vice versa, as the performance that results will be different. {It is not just deciding whether to serve a larger number of users with lower QoS or a smaller number of users with higher QoS, but also prioritizing the tradeoff  between sum-rate and user QoS.} Additionally, the network conditions play a role on performance and may require reevaluating these decisions. {In a similar manner, clustering and determining the cluster objective impact one another. Having selected which users are in a cluster may require altering the objective according to the needs of the users. Similarly, selecting a cluster based on an objective function may affect which users can be considered. For instance, very weak users would be removed from the pool of potential candidates if the minimum QoS constraint required by the objective is too high.}

\subsection*{Resource Allocation}
Resource allocation for NOMA refers to the power given to the messages of the users sharing a resource block. When fixed-rate transmissions are used, deciding the target-rate of each message is also included in the resource allocation process. While it makes sense to allocate resources in a manner that allots higher power and/or lower target rate to the messages of users that need to be decoded first in the SuIC chain (making them easier to decode), it is not compulsory. In fact, the rate region is made up of every possible combination of resources allocated to each user. That said, the importance of careful resource allocation needs to be emphasized. {A condition called the NOMA necessary condition for coverage is introduced in \cite{myNOMA_tcom,myNOMA_icc} which shows that certain combinations of the allocated resources, because of the nature of SuIC, cause guaranteed outage.} Even when this condition is satisfied, careful/smart resource allocation is crucial for improving performance; optimum resource allocation results in operating at the boundary of the rate region. 

A number of works on NOMA in the literature assume fixed resource allocation; however, this results in far from optimum performance. Optimum resource allocation in the context of a large network is a challenging problem to solve. The optimization problem to be solved of course varies with the cluster and network objectives; hence, prior clustering can also impact resource allocation. Often the problems are non-convex and an exhaustive search is required to solve them. A number of works resort to suboptimum solutions for these problems as an exhaustive search can become very tedious very fast as the cluster size grows. Consideration of multiple antenna techniques brings additional complexity to the problem. Resource allocation for NOMA is an open problem for the next generation cellular networks using large number of antennas for radio transmission \cite{N2}.


{It should also be mentioned that while a number of works on NOMA focus on and study one resource block, it is important to realize that clustering, by definition, means deciding which users are selected to operate in a certain resource block. There have been works that include channel assignment selection, in terms of resource block selection, in their resource allocation in a single cell setup \cite{N15_19}. While taking channel assignment into account improves the quality of clustering, it significantly complicates the resource allocation problem; hence, there is a trade-off between complexity and performance.}


\section*{Channel Disparity and User Clustering in NOMA: A Closer Look}



 \subsection*{Existing Literature on the Impact of Channel Disparity on NOMA Clustering}
 
The impact of channel disparity explicitly, particularly in the context of large networks, has not been investigated enough. In this subsection we highlight some of the existing contributions; note that while these works do not explicitly measure performance against channel disparity, they take the first step by comparing more and less disparate clustering. The works in \cite{myNOMA_meta,N18} make an effort to compare the impact of random user selection from all over the cell (i.e. no restriction on clustering users in a cell) against more `selective clustering' in large networks. The selective clustering of \cite{N18} places a minimum channel disparity restriction in terms of the SINR between the selected users. The selective clustering in \cite{myNOMA_meta}, on the other hand, places spatial restriction by guaranteeing that the users are selected from the largest disk centred at the BS that is guaranteed to be inside in the Voronoi cell, thus guaranteeing a certain channel quality of users. It should be noted that the two works emphasize different selective clustering; while \cite{N18} promotes selective clustering of users with higher channel disparity, \cite{myNOMA_meta} promotes selective clustering of users with lower channel disparity but overall better channel conditions. Both works show that their selective clustering outperforms the case where users are selected uniformly at random from all over the cell which is contradictory. It should be noted that \cite{N18} uses fixed power allocation while \cite{myNOMA_meta} uses optimum resource allocation. We believe this contradiction occurs because the work in \cite{N18} does clustering after resource allocation; thus, selecting users with a certain minimum channel disparity in this setup may be superior to selecting users at random without any restriction; however, it cannot be generalized to setups with other fixed resource allocation values or when clustering is done prior to resource allocation. 

In \cite{N19}, two-user downlink NOMA in a single cell setup is considered. Multiple users are available to select the NOMA pair from. The authors show that under fixed transmit powers (F-NOMA), having a higher channel disparity between the two users results in a better performance. They also show that in a cognitive radio inspired NOMA setup (CR-NOMA), where the weak user must attain QoS and the strong user can access the remaining resources, having lower channel disparity is more beneficial. We would like to highlight that this may be the case for F-NOMA because the power allocation is not optimum given the channel conditions of the users. 




\subsection*{On Myths in NOMA Clustering}\label{3a}

There is a common myth (or misunderstanding) that employing NOMA in a manner that allows users with similar channel strengths, i.e. low channel disparity, to share a resource block will be dangerously detrimental for performance. Often, the rationale behind this misconception is that when users with similar channel strengths employ NOMA, the power allocation will be very similar which will lead to very poor SuIC. {To elaborate on this, in the two-user NOMA case, roughly equal power allocation will make it difficult for both the strong and weak user to decode the message of the weak user, due to very large intracell interference, causing them to be in outage.} In fact, a number of works promote NOMA for users that have very large channel disparity between them under the assumption that they will make a `good pair'. However, this may not be necessarily true. {Although having a very high channel disparity leads to very high powers being allocated to the weak user which does make decoding the weak user's message very easy, {a trade off exists for the strong user between having power to decode the weak user's message easily and having power for enhancing its rate.} As we will show, when channel disparity is low, similar power allocation does not become a bottleneck for coverage.} 

NOMA is intended to serve multiple users in a resource block. However, these users need to be strong enough to share a resource block. As such, increasing channel disparity beyond a point makes the weak users too weak defeating the purpose of NOMA which is service of multiple users, including weak users, in this resource block. Under such circumstances, the very weak user is better off employing OMA or even techniques such as coordinated multipoint (CoMP) transmissions. {In terms of sum-rate,  in the two-user case, for example, two strong users will do much better than one strong and one weak user. This is simply because each strong user has a better channel and can therefore exploit any given resources more than a weaker user.} 

\subsection*{Unveiling the Realities With NOMA Clustering: Numerical Examples}


We consider two-user downlink NOMA in a large network where the BSs are distributed according to a Poisson point process. We denote the distance between the serving BS and the nearest interfering BS by $\rho$. The link distance of the strong user is fixed to $\rho/4$ in any random orientation. The weak user's distance is increased from $\rho/4$ in random orientations, while ensuring that the weak user is also inside the cell. Simulations are repeated 100,000 times. Note that in such a setup, weak users with link distance upto $\rho/2$ are guaranteed to be inside the cell as this is the distance to the nearest cell edge from the BS. We measure channel disparity between the two users in terms of the ratio of the link distance of the weak user to the strong user; hence, it increases from 1. A QoS-constrained system is considered and we assume a total power budget of 1. We consider two set ups: the first set up attempts to maximize sum rate; as a result, power allocation is such that the minimum power required to attain QoS for the weak user is allocated and the remaining power is given to the strong user to maximize its rate; and the second set up  only attains QoS for each user using the minimum required power.



\begin{figure}[htb]
\begin{minipage}[htb]{\linewidth}
\centering\includegraphics[width=0.625\columnwidth]{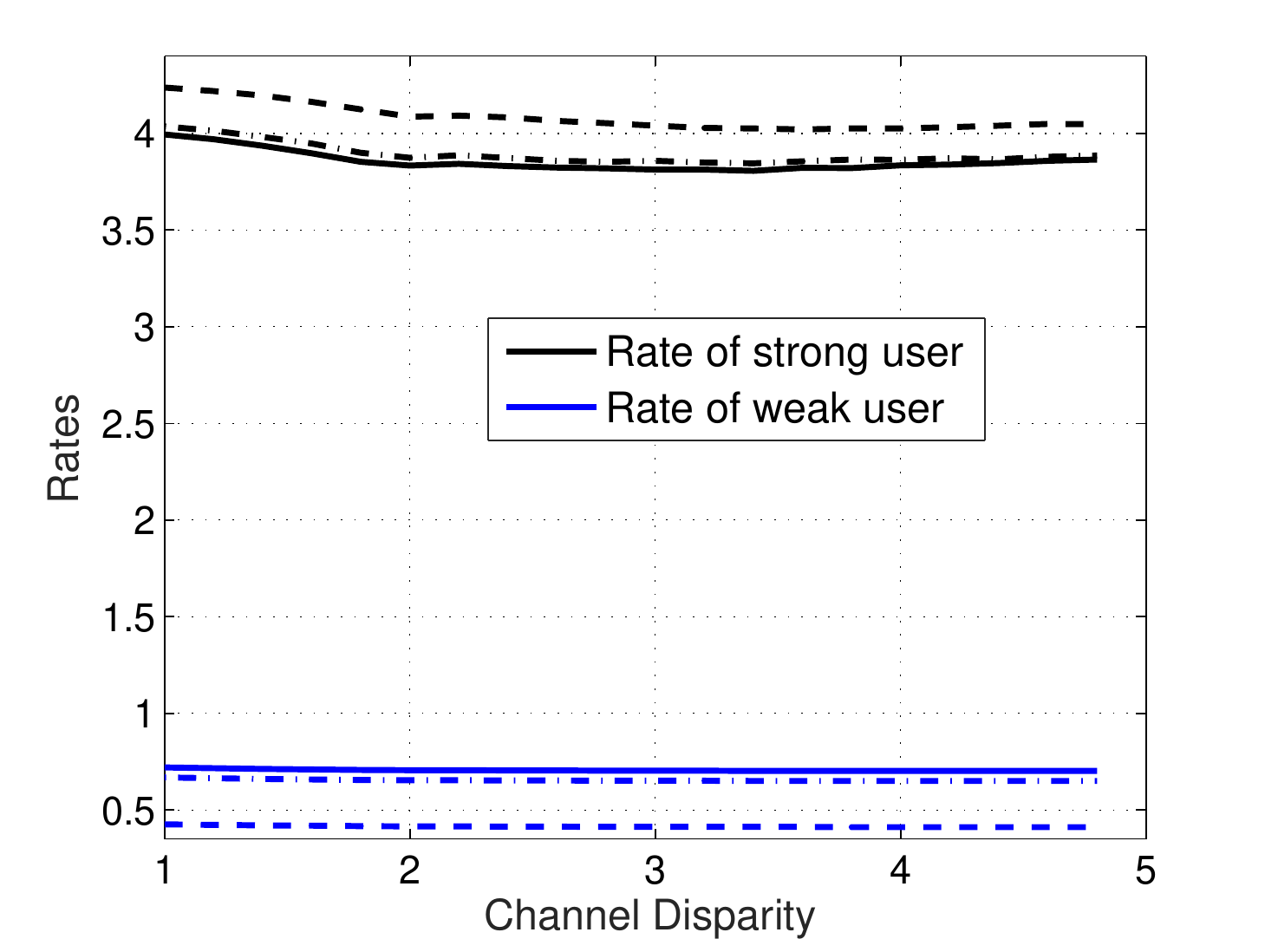}
\subcaption{User rates against increasing channel disparity.}\label{rateVsDisp}
\end{minipage}
\begin{minipage}[htb]{\linewidth}
\centering\includegraphics[width=0.625\columnwidth]{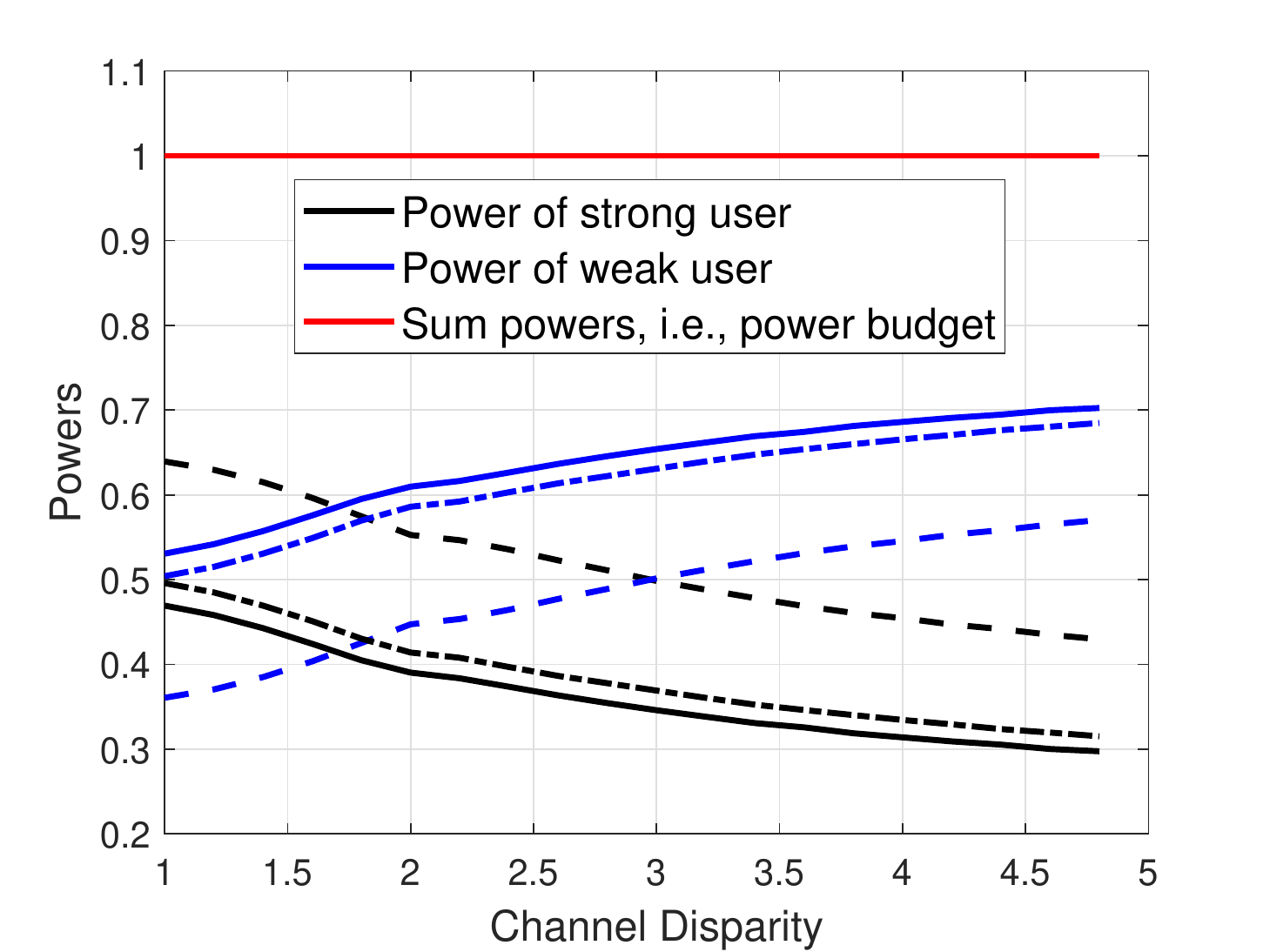}
\subcaption{Powers against increasing channel disparity.}\label{powerVsDisp}
\end{minipage}
\begin{minipage}[htb]{\linewidth}
\centering\includegraphics[width=0.625\columnwidth]{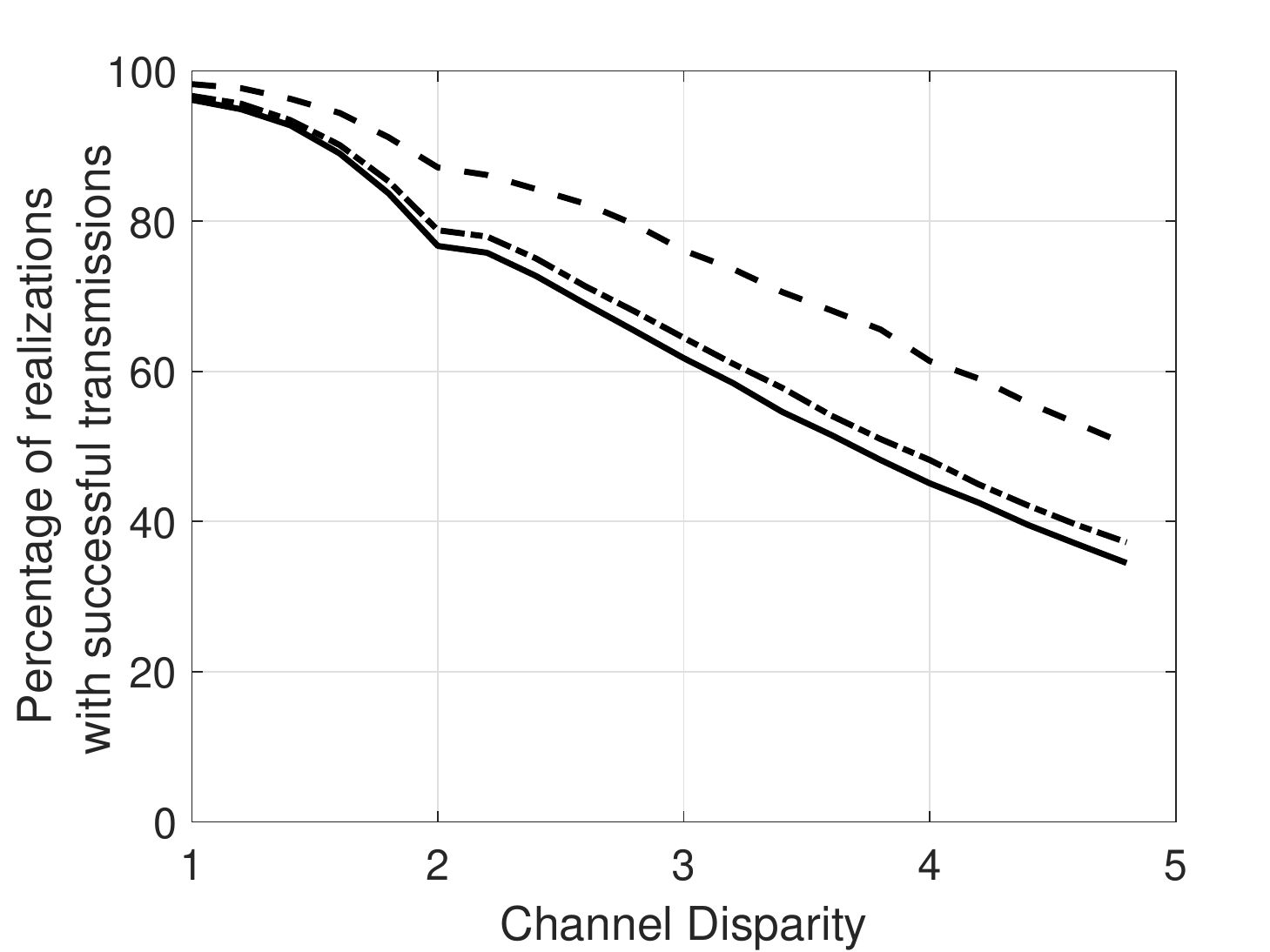}
\subcaption{Percentage of cells that can attain a size corresponding to the channel disparity.}\label{perSimsVsDisp}
\end{minipage}
\caption{A two-user downlink NOMA setup. The system maximizes sum rate subject to a QoS constraint of $\log(1+\theta)$ and power budget of 1. QoS constraints corresponding to $\theta$ values of 0.5, 0.9, and 1 are compared using dashed, dash-dotted, and solid lines, respectively.}\label{rateNpowerVsDisp}
\end{figure}

{For the first set up, Figs. \ref{rateVsDisp}, \ref{powerVsDisp}, and \ref{perSimsVsDisp} plot the rates attained by the users, the powers allocated to the users, and the percentage of realizations that have successful transmissions with increasing channel disparity, respectively, for three different QoS requirements. As anticipated, we observe in Fig. \ref{powerVsDisp} that the power allocated to the weak user increases with channel disparity, and accordingly the fraction of power left for the strong user decreases with disparity. A number of important observations and insights are listed below:

 \begin{itemize}
 \item The rate of the strong user decreases as channel disparity increases at first since less power is left for the strong user. This negates the proposition that increasing channel disparity always improves performance and is therefore always beneficial to NOMA.
 
 \item The maximum performance in terms of highest rate of the strong user is achieved when channel disparity is at its lowest, i.e., the two users have very similar channel conditions. 

\item When QoS is $\log(1+0.9)$, the lowest disparity also results in equivalent power allocation and we still observe maximum performance showing that equal power allocation when the two users have very low channel disparity is not detrimental to performance.

\item The percentage of realizations that have successful transmissions decreases significantly with channel disparity. In the considered setups, channel disparity is conditioned on link distances; when it is low, link distances of the users are already good so transmission failures are due to variables such as fading\footnote{This is also why when channel disparity is between 1 and 2, where both users are always guaranteed to be in the cell, 100\% successful transmissions are not observed in Fig. \ref{perSimsVsDisp}, highlighting the impact of deep fades.}. However, at high channel disparity, link distances are already bad so successful transmissions occur only when fading is very good.

\item Performance does not decrease monotonically with channel disparity; in fact, when channel disparity is very high, the rate of the strong user increases again. This is because high disparity corresponds to the fewer realizations with very good fading conditions and therefore channel quality; the transmissions that occur during these good conditions therefore result in better rates. 

\item Although high channel disparity may not result in the worst rates, the percentage of successful transmissions under these conditions is very low; the effective rates under such circumstances may thus be very low.

\item The percentage of successful transmissions decreases as QoS increases and at higher QoS, the rate of decrease is also higher. This is because attaining a higher QoS is more difficult, making communication more sensitive to bad fading conditions and therefore more prone to outage.  

\end{itemize}  } 

\begin{figure}[htb]
\begin{minipage}[htb]{\linewidth}
\centering\includegraphics[width=0.625\columnwidth]{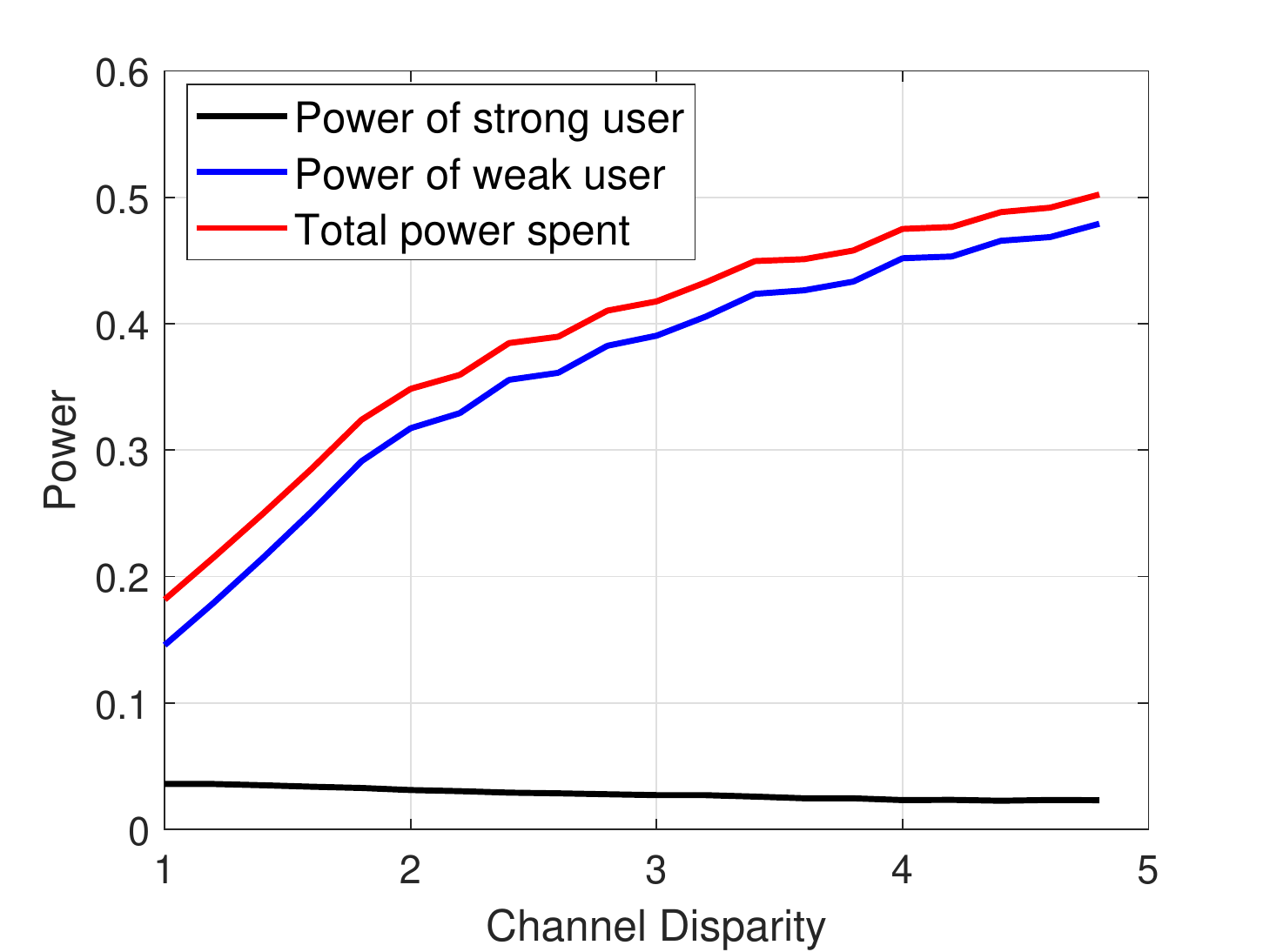}
\subcaption{Powers against increasing channel disparity.}\label{powerVsDisp_allQoS}
\end{minipage}
\begin{minipage}[htb]{\linewidth}
\centering\includegraphics[width=0.625\columnwidth]{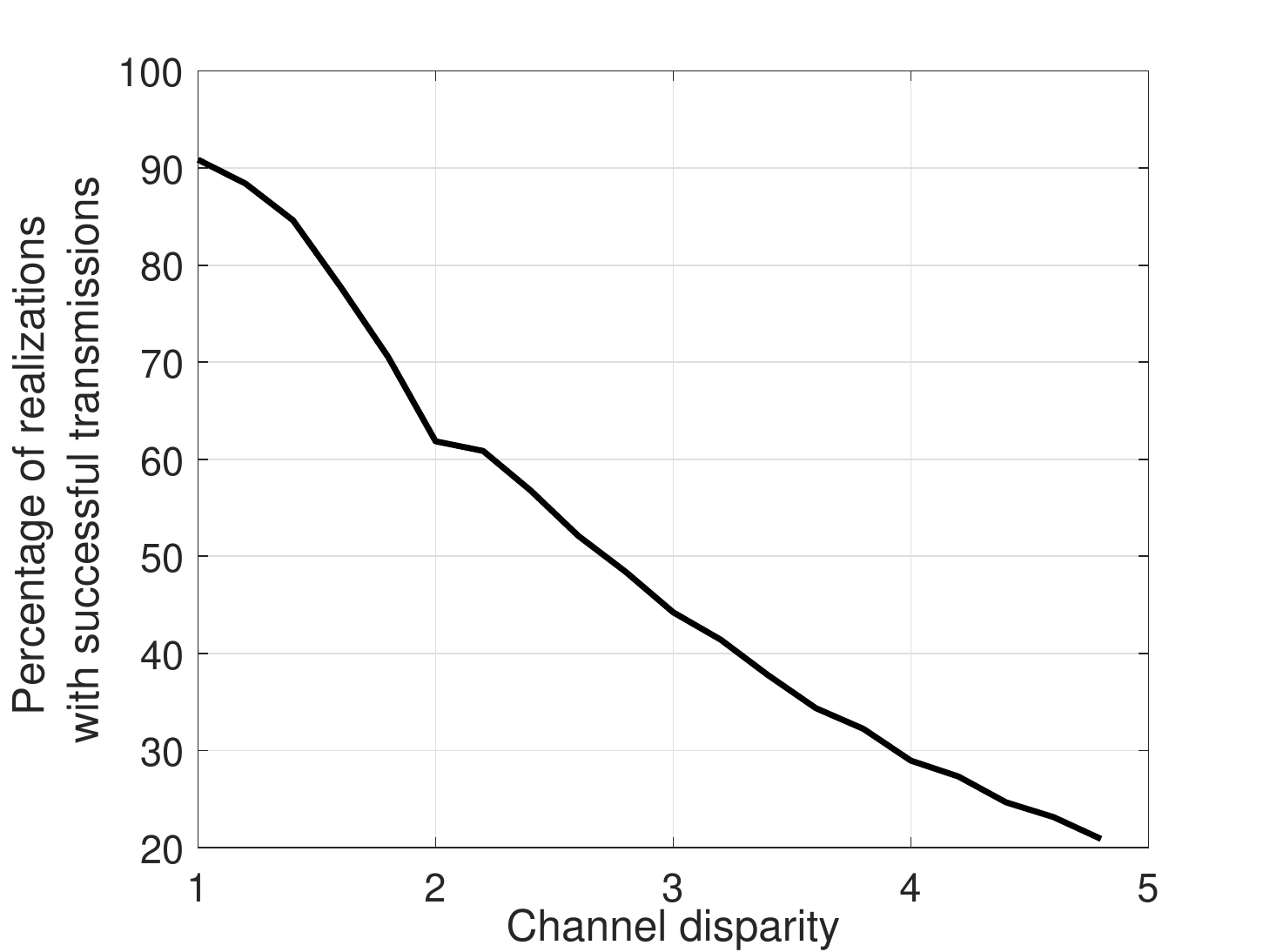}
\subcaption{Percentage of cells that can attain a size corresponding to the channel disparity.}\label{perSimsVsDisp_allQoS}
\end{minipage}
\caption{A two-user downlink NOMA setup. The system attains the QoS constraint of $\log(1+2)$ for each user without exceeding the power budget of 1.}\label{rateNpowerVsDisp_allQoS}
\end{figure}

 In Fig. \ref{powerVsDisp_allQoS} and \ref{perSimsVsDisp_allQoS}, we plot the powers allocated to the users and percentage of realizations that have successful transmissions as channel disparity increases, respectively, for the second setup. We do not plot the attained rates of the users as each simply attains the QoS. The rationale behind plotting Fig. \ref{powerVsDisp_allQoS} is to show what fraction of power the strong user requires in order to attain just the QoS. This is unlike Fig. \ref{rateNpowerVsDisp} where it uses all of the remaining power to maximize its rate. While one would anticipate that the power required by the strong user should be constant as its link distance is fixed and its performance should therefore not be affected by channel disparity, we observe that this is not the case.  The power of the strong user decreases with channel disparity. This again occurs because increasing channel disparity corresponds to successful transmissions that have improving fading conditions; lower power is therefore required by the strong user to attain QoS. The weak user's required power, as expected, grows with increasing channel disparity because factors such as link distance are still the bottleneck for attaining QoS even though fading conditions are better in the successful transmissions at high channel disparity. It should also be noted that in this setup, the goal is to obtain the same QoS for the two users. Hence, selecting NOMA users that have lower channel disparity to share a resource block leaves behind more resources (i.e. more power) to either spend on improving the performance of one or both of the users, or to add additional users to the system.
 

It should also be highlighted that since intercell interference impacts the SINR and therefore the quality of the user, the disparity between users reduces substantially with increasing traffic. This in turn affects resource allocation and performance of  all the users. The impact of intercell interference is shown explicitly in Fig. 5 of \cite{my_nomaMag} where a notable reduction in the discrepancy between power allocated to each user is observed when intercell interference is considered versus when it is not. This implies that the channel conditions of the users become similar in high traffic scenarios. Using a more discrepant power allocation in such a situation would deteriorate performance drastically as shown in Fig. 1 of \cite{my_nomaMag}; thus highlighting again that having similar power allocation does not mean that performance will deteriorate. 



\subsection*{On Employing NOMA When Channel Disparity is High}
Employing NOMA for users with high channel disparity can still be desirable if the goal is to help weaker users achieve better rates via transmissions, albeit shared, over multiple resource blocks. However,  this is to improve a goal such as the sum-rate of very weak users over \emph{multiple} resource blocks; in one resource block, such users are better off employing OMA. Employing NOMA in such a manner may be of interest to maximize efficiency given the type of users available.


Additionally, employing NOMA for users with high channel disparity is also of interest when QoS is flexible and there are enough power resources. In such a situation, the target rates for weak users can be chosen accordingly so that they do not become the bottleneck for power consumption. In this way, the weak users with low target rates can be clustered with much stronger users. With the very wide variety of IoT devices that are to be deployed in next generation networks, NOMA would therefore be useful. 


\section*{Conclusion}
We have given a brief overview of multiple access techniques in cellular systems while focusing on NOMA. We have discussed important challenges associated with NOMA emphasizing on the clustering problem. We have examined how the challenges are intertwined and particularly emphasized the impact of clustering on each. We have highlighted some open problems in this context. We have reviewed the existing literature on exploiting channel disparity in NOMA clustering. {We have shown through numerical illustrations that the misunderstandings that: 1) clustering users with low channel disparity is detrimental, 2) similar power allocation is detrimental for NOMA, are incorrect.}


%
%

\appendices

\bibliographystyle{IEEEtran}
\bibliography{refsNOMA}

\begin{thebibliography}{10}
\providecommand{\url}[1]{#1}
\csname url@samestyle\endcsname
\providecommand{\newblock}{\relax}
\providecommand{\bibinfo}[2]{#2}
\providecommand{\BIBentrySTDinterwordspacing}{\spaceskip=0pt\relax}
\providecommand{\BIBentryALTinterwordstretchfactor}{4}
\providecommand{\BIBentryALTinterwordspacing}{\spaceskip=\fontdimen2\font plus
\BIBentryALTinterwordstretchfactor\fontdimen3\font minus
  \fontdimen4\font\relax}
\providecommand{\BIBforeignlanguage}[2]{{%
\expandafter\ifx\csname l@#1\endcsname\relax
\typeout{** WARNING: IEEEtran.bst: No hyphenation pattern has been}%
\typeout{** loaded for the language `#1'. Using the pattern for}%
\typeout{** the default language instead.}%
\else
\language=\csname l@#1\endcsname
\fi
#2}}
\providecommand{\BIBdecl}{\relax}
\BIBdecl

\bibitem{N1}
Y.~Saito, Y.~Kishiyama, A.~Benjebbour, T.~Nakamura, A.~Li, and K.~Higuchi,
  ``Non-orthogonal multiple access ({NOMA}) for cellular future radio access,''
  in \emph{Proc. of IEEE 77th Vehicular Technology Conference (VTC Spring
  2013)}, June 2013, pp. 1--5.

\bibitem{N8}
Z.~Ding, Z.~Yang, P.~Fan, and H.~V. Poor, ``On the performance of
  non-orthogonal multiple access in {5G} systems with randomly deployed
  users,'' \emph{IEEE Signal Proc. Letters}, vol.~21, no.~12, pp. 1501--1505,
  Dec. 2014.

\bibitem{myNOMA_icc}
K.~S. Ali, H.~ElSawy, A.~Chaaban, M.~Haenggi, and M.~Alouini, ``Analyzing
  non-orthogonal multiple access ({NOMA}) in downlink {P}oisson cellular
  networks,'' in \emph{Proc. of IEEE International Conference on Communications
  (ICC18)}, May 2018, pp. 1--6.

\bibitem{myNOMA_tcom}
K.~S. Ali, M.~Haenggi, H.~E. Sawy, A.~Chaaban, and M.~Alouini, ``Downlink
  non-orthogonal multiple access {(NOMA)} in {P}oisson networks,'' \emph{IEEE
  Trans. Commun.}, vol.~67, no.~2, pp. 1613--1628, Feb. 2019.

\bibitem{N2}
Y.~Liu, Z.~Qin, M.~Elkashlan, Y.~Gao, and A.~Nallanathan, ``Non-orthogonal
  multiple access in massive {MIMO} aided heterogeneous networks,'' in
  \emph{Proc. of IEEE Global Communications Conference (GLOBECOM16)}, Dec.
  2016.

\bibitem{N9}
S.~Timotheou and I.~Krikidis, ``Fairness for non-orthogonal multiple access in
  {5G} systems,'' \emph{IEEE Signal Proc. Letters}, vol.~22, no.~10, pp.
  1647--1651, Oct. 2015.

\bibitem{N15}
J.~Zhu, J.~Wang, Y.~Huang, S.~He, X.~You, and L.~Yang, ``On optimal power
  allocation for downlink non-orthogonal multiple access systems,'' \emph{IEEE
  J. Selec. Areas Commun.}, vol.~35, no.~12, pp. 2744--2757, Dec. 2017.

\bibitem{my_nomaMag}
K.~S. Ali, H.~Elsawy, A.~Chaaban, and M.~S. Alouini, ``Non-orthogonal multiple
  access for large-scale {5G} networks: Interference aware design,'' \emph{IEEE
  Access}, vol.~5, pp. 21\,204--21\,216, 2017.

\bibitem{N6}
H.~Tabassum, E.~Hossain, and M.~J. Hossain, ``Modeling and analysis of uplink
  non-orthogonal multiple access ({NOMA}) in large-scale cellular networks
  using {P}oisson cluster processes,'' \emph{IEEE Trans. Commun.}, vol.~65,
  no.~8, pp. 3555--3570, Aug. 2017.

\bibitem{N18}
Z.~Zhang, H.~Sun, and R.~Q. Hu, ``Downlink and uplink non-orthogonal multiple
  access in a dense wireless network,'' \emph{IEEE J. Selec. Areas Commun.},
  vol.~35, no.~12, pp. 2771--2784, Dec. 2017.

\bibitem{myNOMA_meta}
K.~S. Ali, H.~E. Sawy, and M.~Alouini, ``Meta distribution of downlink
  non-orthogonal multiple access ({NOMA}) in {P}oisson networks,'' \emph{IEEE
  Wireless Comm. Letters}, vol.~8, no.~2, pp. 572--575, Apr. 2019.

\bibitem{N16}
Z.~Zhang, H.~Sun, R.~Q. Hu, and Y.~Qian, ``Stochastic geometry based
  performance study on {5G} non-orthogonal multiple access scheme,'' in
  \emph{Proc. of IEEE Global Communications Conference (GLOBECOM16)}, Dec.
  2016, pp. 1--6.

\bibitem{N4}
Y.~Liu, Z.~Ding, M.~Elkashlan, and H.~V. Poor, ``Cooperative non-orthogonal
  multiple access with simultaneous wireless information and power transfer,''
  \emph{IEEE J. Select. Areas Commun.}, vol.~34, no.~4, pp. 938--953, Apr.
  2016.

\bibitem{N15_19}
Y.~Sun, D.~W.~K. Ng, Z.~Ding, and R.~Schober, ``Optimal joint power and
  subcarrier allocation for full-duplex multicarrier non-orthogonal multiple
  access systems,'' \emph{IEEE Trans. Commun.}, vol.~65, no.~3, pp. 1077--1091,
  Mar. 2017.

\bibitem{N19}
Z.~Ding, P.~Fan, and H.~V. Poor, ``Impact of user pairing on {5G} nonorthogonal
  multiple-access downlink transmissions,'' \emph{IEEE Trans. Vehicular Tech.},
  vol.~65, no.~8, pp. 6010--6023, Aug. 2016.

\end{thebibliography}

\end{document}